\documentclass[prd,11pt,nofootinbib]{revtex4}

\usepackage[force]{feynmp-auto} 
\usepackage{amsmath}    
\usepackage{graphicx}   
\usepackage{subcaption} 
\usepackage{verbatim}   
\usepackage{color}      
\usepackage{hyperref}   
\usepackage{slashed}
\usepackage{adjustbox}  

\newcommand{\be}{\begin{equation}}
\newcommand{\ee}{\end{equation}}
\newcommand{\bea}{\begin{eqnarray}}
\newcommand{\eea}{\end{eqnarray}}
\newcommand{\nn}{\nonumber}

\begin{document}
 
 \begin{flushright}
 CP3-19-03
\end{flushright}

\title{New Physics in Double Higgs Production at Future $e^+ e^-$ Colliders  }
\author{Andres Vasquez$^{1,2}$}
\author{C\'eline Degrande$^{2}$}
\author{Alberto Tonero$^3$} 
\author{Rogerio Rosenfeld$^1$}
\affiliation{$^1$ICTP South American Institute for Fundamental Research \& 
Instituto de F\'{\i}sica Te\'orica \\
UNESP - Universidade Estadual Paulista \\
Rua Dr.~Bento T.~Ferraz 271  -  01140-070  S\~ao Paulo, SP, Brazil}
\affiliation{$^2$Centre for Cosmology, Particle Physics and Phenomenology (CP3), Universit\'e Catholique de Louvain, B-1348 Louvain-la-Neuve, Belgium\\}
\affiliation{$^3$Ottawa-Carleton, Institute for Physics, Carleton University
1125 Colonel By Drive, Ottawa, ON, K1S 5B6, Canada\\}
\date{\today}

\begin{abstract}
We study the effects of new physics in double Higgs production at future $e^+ e^-$ colliders. In the Standard Model the chiral limit ($m_e=0$) plays an important role for this process, being responsible for the smallness of the tree-level diagrams with respect to the 1-loop contributions. In our work, we consider the possibility of an enhancement due to the contribution of Standard Model dimension-six effective operators.
We show that there are only two relevant operators for this process that are not yet (strongly) constrained by other data. We perform a sensitivity study on the operator coefficients for several benchmark values of energy and integrated luminosity related to the proposed linear colliders such as CLIC, ILC and FCC-ee and we derive expected 95\% CL limits for each benchmark scenario.
\end{abstract}

\maketitle

\section{Introduction}
The discovery of the Higgs boson during the Run I of the LHC~\cite{Chatrchyan:2012ufa} has put in place the main building block that was missing for the experimental validation of the Standard Model (SM). Since then, a great effort has been made by the experimental collaborations in the attempt to define the properties of this new particle, namely its mass, spin, parity and coupling to itself and the other particles of the SM, mainly through global fits in the so-called kappa framework~\citep{Khachatryan:2016vau,Aaboud:2018xdt}.

These analyses are crucial to pin down the Higgs boson properties and to understand the nature of electroweak symmetry breaking (EWSB). These studies will play a fundamental role especially during the high-luminosity run of the LHC (HL-LHC) as well as for the future hadron and lepton colliders. Any deviations from SM predictions would unravel the presence of new physics. 

As the second run of the LHC is coming to an end, no clear signs of new physics have been found yet. This fact points to a scenario in which new physics is most probably out of the reach of the LHC and in this case the best way to search for it is through indirect effects via precision measurements. 

Precision studies of the properties of the Higgs boson and the nature of the electroweak symmetry breaking strongly motivate the construction of a lepton collider which benefits from a cleaner environment with respect to hadron colliders. There have been several proposals for a future electron-positron collider, such as the Compact Linear Collider (CLIC)~\cite{CLIC}, the International Linear Collider (ILC) \cite{ILC}, the Circular Electron Positron Collider (CPEC)~\cite{CPEC} and the Future Circular Collider with $e^+ e^-$ (FCC-ee) at CERN, previously known as TLEP~\cite{fcc-ee}.

The main production mechanism of the Higgs boson at $e^+e^-$ colliders is the bremsstrahlung process $e^+e^-\to h Z$ (Higgsstrahlung). At a center-of-mass energies of 240-250 GeV, close to the maximum of the Higgsstrahlung cross section, this process will allow to determine Higgs couplings to gauge bosons with unprecedented precision. In addition there are also weak boson fusion production processes $e^+e^-\to W^*W^*/Z^*Z^*\to h \nu \bar\nu/ h e^+e^-$ which provide an increasingly powerful handle at higher center-of-mass energies. Finally, also the process $e^+ e^- \to tth$ benefits from high energies and represent an important measurement to directly constrain the top Yukawa coupling. A comprehensive sensitivity study about the effect of new physics, parametrized by higher-dimensional operators, affecting these production mechanisms for the different proposed $e^+e^-$ machines have been performed in~\cite{Ellis:2017kfi,Durieux:2017rsg,DiVita:2017vrr}.

Some of the Higgs boson couplings can also be tested in higher order processes involving for instance Higgs pair production. In this case, the Higgs self coupling and the couplings to gauge bosons can be measured in the so-called double higgstrahlung ($e^+ e^- \rightarrow hhZ$) and vector boson fusion ($e^+ e^- \rightarrow e^+ e^- (\nu \bar{\nu}) hh$) processes~\cite{Boudjema:1995cb,Haba:2013xla,Panico:2015cst}. The top Yukawa can be measured in double Higgs production in association with top quarks ($e^+ e^- \rightarrow hh t \bar{t}$). These processes are tree-level dominated processes but compared to the previous ones they are characterized by higher orders in the coupling constants.

On the other hand, the process $e^+ e^- \rightarrow hh$, where only two Higgs bosons are actually produced in the final state, is completely dominated by the contribution of one-loop diagrams and therefore one can test higher order effects in a clean way because they are not masked by tree diagram contributions. For instance, it can be useful to discriminate between the Higgs sector of the Standard Model from the more complicated scalar sectors belonging to possible extensions, {\it e.g.} two Higgs doublet model~\cite{Djouadi:1996hp,LopezVal:2009qy}.

At hadron colliders, double Higgs production via gluon fusion at LHC has been exhaustively studied as a probe of physics beyond the SM~\cite{hh-pp}. The sensitivity to new physics is enhanced due to a cancellation between triangle and box contributions in the gluon fusion process in the SM \cite{cancellation}. It is well known that Higgs pair production at hadron colliders is sensitive to new physics effects parametrized by higher-dimensional operators~\cite{Goertz:2014qta,effective}. On the other hand, an enhancement in the cross section can also arise from the presence of an hidden sector, as studied in \cite{Oliveira:2010uv}.
Double Higgs production has also been studied as probe for Higgs anomalous couplings at future electron-proton colliders~\cite{Kumar:2015kca}.

The SM cross section for double Higgs production at the LHC is not very large (approximately 37 fb at 14 TeV at NNLO) and the background can be challenging even for HL-LHC. Therefore, the cleaner environment of an electron-positron collider could be very helpful to find deviations from the SM or to improve bounds on new physics. 

In this work we will proceed in that direction and focus on the process  $e^+ e^- \rightarrow hh$ at future lepton colliders as a probe of new physics which we take to be parametrized by the presence of dimension-six effective operators of the SM effective field theory (SMEFT).

This paper is organized as follows: in section II we revise the SM computation for the process  $e^+ e^- \rightarrow hh$. In section III we highlight the relevant SMEFT contributions that we consider in our study. In section IV we discuss different benchmark scenarios for future  $e^+ e^-$ colliders, we present the analysis strategy and we report the 95\% CL bounds on the operator coefficients for each benchmark scenario. In section V we conclude.

\section{SM double Higgs production at $e^+e^-$ colliders}
\begin{figure}[ht!]
\centering
\unitlength = 1mm
\begin{subfigure}{0.3\textwidth}
\begin{fmffile}{TrLev1}
\begin{fmfgraph*}(40,20)
\fmfset{thin}{0.8pt}
\fmfset{arrow_len}{2.5mm}
\fmfleft{i1,i2}
\fmfright{o1,o2}
\fmf{fermion}{i1,v1,i2}
\fmf{dashes}{o1,v2,o2}
\fmf{dashes,label=$h$,tension=1.3}{v1,v2}
\fmflabel{$e^-$}{i1}
\fmflabel{$e^+$}{i2}
\fmflabel{$h$}{o1}
\fmflabel{$h$}{o2}
\end{fmfgraph*}
\end{fmffile}
\label{TrLev1}
\caption{ }
\end{subfigure}
\qquad
\begin{subfigure}{0.3\textwidth}
\begin{fmffile}{TrLev2}
\begin{fmfgraph*}(40,20)
\fmfset{thin}{0.8pt}
\fmfset{arrow_len}{2.5mm}
\fmfleft{i1,i2}
\fmfright{o1,o2}
\fmf{fermion}{i1,v1}
\fmf{dashes}{v1,o1}
\fmf{fermion}{v2,i2}
\fmf{dashes}{v2,o2}
\fmf{fermion,tension=0.5}{v1,v2}
\fmflabel{$e^-$}{i1}
\fmflabel{$e^+$}{i2}
\fmflabel{$h$}{o1}
\fmflabel{$h$}{o2}
\end{fmfgraph*}
\end{fmffile}
\label{TrLev2}
\caption{ }
\end{subfigure}
\caption{SM tree level diagrams for $e^+ e^- \to hh$.}
\label{smtree}
\end{figure}
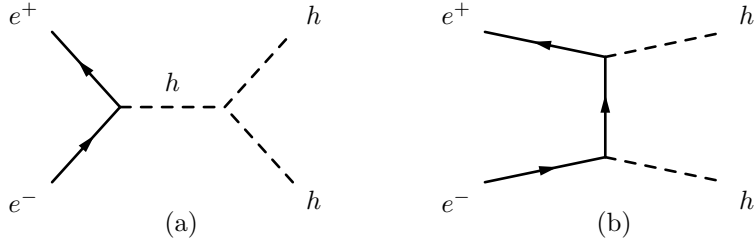

The process $e^+ e^- \to hh$ is an interesting one from the theoretical point of view because SM tree level diagrams (see Fig.~ \ref{smtree}) give a negligible contribution to the cross section since they are proportional to $m_e/\upsilon$, where $m_e$ is the electron mass and $\upsilon=246$ GeV is the Higgs vacuum expectation value (VEV). This fact has been recognized long ago and as a consequence the cross section is quite small both in the SM and MSSM extensions~\cite{Gaemers:1984vw, Djouadi:1996hp,LopezVillarejo:2008xw}.

\begin{figure}[ht!]
\centering
\unitlength = 1.2mm
\begin{subfigure}{0.3\textwidth}
\begin{fmffile}{Loop1}
\begin{fmfgraph*}(40,20)
\fmfset{thin}{0.8pt}
\fmfset{arrow_len}{2.5mm}
\fmfleft{i1,i2}
\fmfright{o1,o2}
\fmf{fermion,tension=1.2}{i1,v1}
\fmf{photon,label=$Z,,W$,l.side=right}{v1,v3}
\fmf{fermion,tension=1.2}{v2,i2}
\fmf{photon,label=$Z,,W$,l.side=left}{v2,v4}
\fmf{fermion,label=$e,,\nu$,l.side=left,tension=0}{v1,v2}
\fmf{dashes,tension=1}{v3,o1}
\fmf{dashes,tension=1}{v4,o2}
\fmf{fermion,tension=40}{v3,v4}
\fmflabel{$e^-$}{i1}
\fmflabel{$e^+$}{i2}
\fmflabel{$h$}{o1}
\fmflabel{$h$}{o2}
\end{fmfgraph*}
\end{fmffile}
\label{Loop1}
\caption{ }
\end{subfigure}
\qquad
\begin{subfigure}{0.3\textwidth}
\begin{fmffile}{Loop2}
\begin{fmfgraph*}(40,20)
\fmfset{thin}{0.8pt}
\fmfset{arrow_len}{2.5mm}
\fmfleft{o1,o2}
\fmfright{i1,i2}
\fmf{dashes,tension=1.4}{i1,v1}
\fmf{photon,label=$W$,l.side=left}{v1,v3}
\fmf{dashes,tension=1.4}{v2,i2}
\fmf{photon,label=$W$,l.side=right}{v2,v4}
\fmf{boson,label=$W$,l.side=right,tension=-0.2}{v1,v2}
\fmf{boson,label=$\gamma,,Z$,tension=1.5}{v5,v4}
\fmf{fermion,tension=1.5}{o1,v5,o2}
\fmf{photon,tension=40}{v3,v4}
\fmflabel{$e^-$}{o1}
\fmflabel{$e^+$}{o2}
\fmflabel{$h$}{i1}
\fmflabel{$h$}{i2}
\end{fmfgraph*}
\end{fmffile}
\label{Loop2}
\caption{ }
\end{subfigure}
\quad
\begin{subfigure}{0.3\textwidth}
\begin{fmffile}{Loop3}
\begin{fmfgraph*}(40,20)
\fmfset{thin}{0.8pt}
\fmfset{arrow_len}{2.5mm}
\fmfleft{i1,i2}
\fmfright{o1,o2}
\fmf{fermion,tension=1.4}{i1,v1}
\fmf{photon,label=$Z$,l.side=right}{v1,v3}
\fmf{fermion,tension=1.4}{v2,i2}
\fmf{photon,label=$Z$,l.side=left}{v2,v4}
\fmf{fermion,label=$e$,l.side=left,tension=-0.2}{v1,v2}
\fmf{dashes,label=$h$,tension=1.5}{v5,v4}
\fmf{dashes,tension=1.5}{o1,v5,o2}
\fmf{photon,tension=40}{v3,v4}
\fmflabel{$e^-$}{i1}
\fmflabel{$e^+$}{i2}
\fmflabel{$h$}{o1}
\fmflabel{$h$}{o2}
\end{fmfgraph*}
\end{fmffile}
\label{Loop3}
\caption{ }
\end{subfigure}
\caption{SM 1-loop triangle diagrams for $e^+ e^- \to hh$.}
\label{smloop1}
\end{figure}
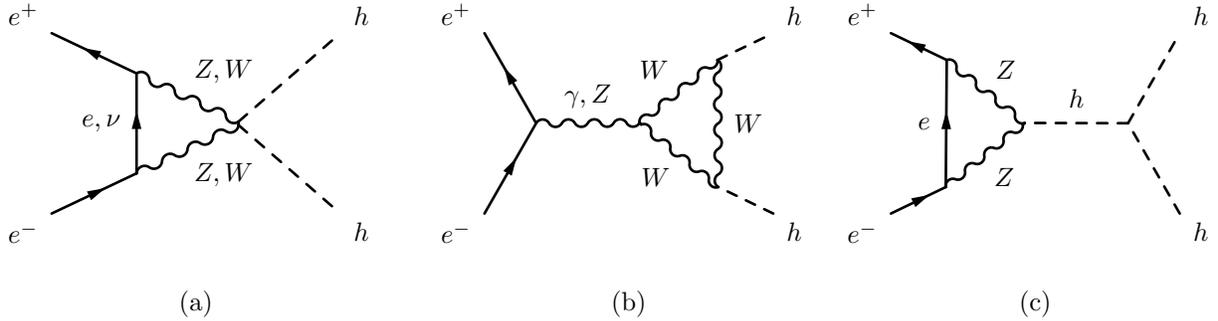

\begin{figure}[ht!]
\vspace{0.5cm}
\centering
\unitlength = 1.2mm
\begin{subfigure}{0.3\textwidth}
\begin{fmffile}{Box1}
\begin{fmfgraph*}(40,20)
\fmfset{thin}{0.8pt}
\fmfset{arrow_len}{2.5mm}
\fmfleft{i1,i2}
\fmfright{o1,o2}
\fmf{fermion}{i1,v1}
\fmf{photon,label=$Z$,l.side=right}{v1,v3}
\fmf{fermion}{v2,i2}
\fmf{photon,label=$Z$,l.side=left}{v2,v4}
\fmf{fermion,label=$e$,l.side=left,tension=0.25}{v1,v2}
\fmf{dashes,tension=1.2}{v3,o1}
\fmf{dashes,tension=1.2}{v4,o2}
\fmf{boson,tension=0.25,label=$Z$,l.side=right}{v3,v4}
\fmflabel{$e^-$}{i1}
\fmflabel{$e^+$}{i2}
\fmflabel{$h$}{o1}
\fmflabel{$h$}{o2}
\end{fmfgraph*}
\end{fmffile}
\label{Box1}
\caption{ }
\end{subfigure}
\qquad
\begin{subfigure}{0.3\textwidth}
\begin{fmffile}{Box2}
\begin{fmfgraph*}(40,20)
\fmfset{thin}{0.8pt}
\fmfset{arrow_len}{2.5mm}
\fmfleft{i1,i2}
\fmfright{o1,o2}
\fmf{fermion}{i1,v1}
\fmf{photon,label=$W$,l.side=right}{v1,v3}
\fmf{fermion}{v2,i2}
\fmf{photon,label=$W$,l.side=left}{v2,v4}
\fmf{fermion,label=$\nu$,l.side=left,tension=0.25}{v1,v2}
\fmf{dashes,tension=1.2}{v3,o1}
\fmf{dashes,tension=1.2}{v4,o2}
\fmf{boson,tension=0.25,label=$W$,l.side=right}{v3,v4}
\fmflabel{$e^-$}{i1}
\fmflabel{$e^+$}{i2}
\fmflabel{$h$}{o1}
\fmflabel{$h$}{o2}
\end{fmfgraph*}
\end{fmffile}
\label{Box2}T
\caption{ }
\end{subfigure}
\caption{SM 1-loop loop box diagrams for $e^+ e^- \to hh$.}
\label{smloop2}
\end{figure}
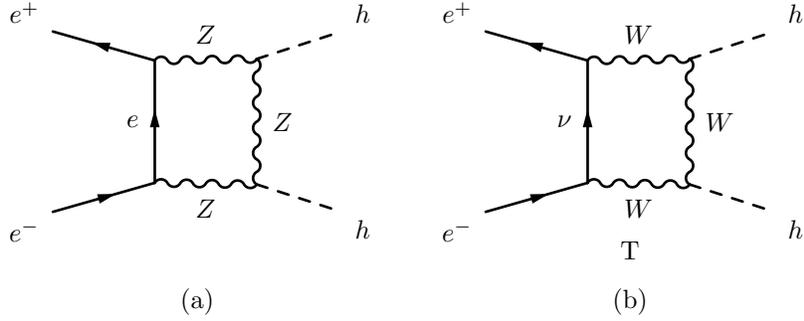

Non-negligible contributions to $e^+ e^- \to hh$ can therefore only come from one-loop diagrams.
In the SM, all one-loop diagrams involving the $\bar e eh$ vertex must give zero contributions in the chiral limit $m_e=0$, to all orders in perturbation theory. Furthermore, because of CP invariance, the diagrams containing intermediate $\gamma$ and $Z$ boson which give rise to two Higgs bosons, also vanish (see  Fig.~\ref{smloop1} (b)). Additional contributions from triangle diagrams involving the quartic $W^+W^-hh/ZZhh$  and triple $hhh$ couplings are also related to the renormalization of the $\bar e eh$ vertex (when one Higgs is taken to its vev) and hence  negligible (see  Fig.~\ref{smloop1} (a) and (c)).  
\newline
\newline
Therefore, the only contribution to Higgs pair production in the SM comes from W and Z box diagrams of Fig.~\ref{smloop2}. Notice that, contrary to double Higgs production in gluon fusion $gg\to hh$, there is no such feature as the cancellation between triangle and box diagrams because the triangle ones are subleading and vanish in the $m_e=0$ limit. Moreover, the dependence of the SM cross section on the triple Higgs coupling $\lambda$ is also negligible because it enters only in diagrams that vanish in the $m_e=0$ limit (see  Fig.~\ref{smloop1} (c)) where the triangle loop is related to the renormalization of the $\bar{e} e h$ coupling.

\begin{figure}
\centering \adjincludegraphics[scale=0.7]{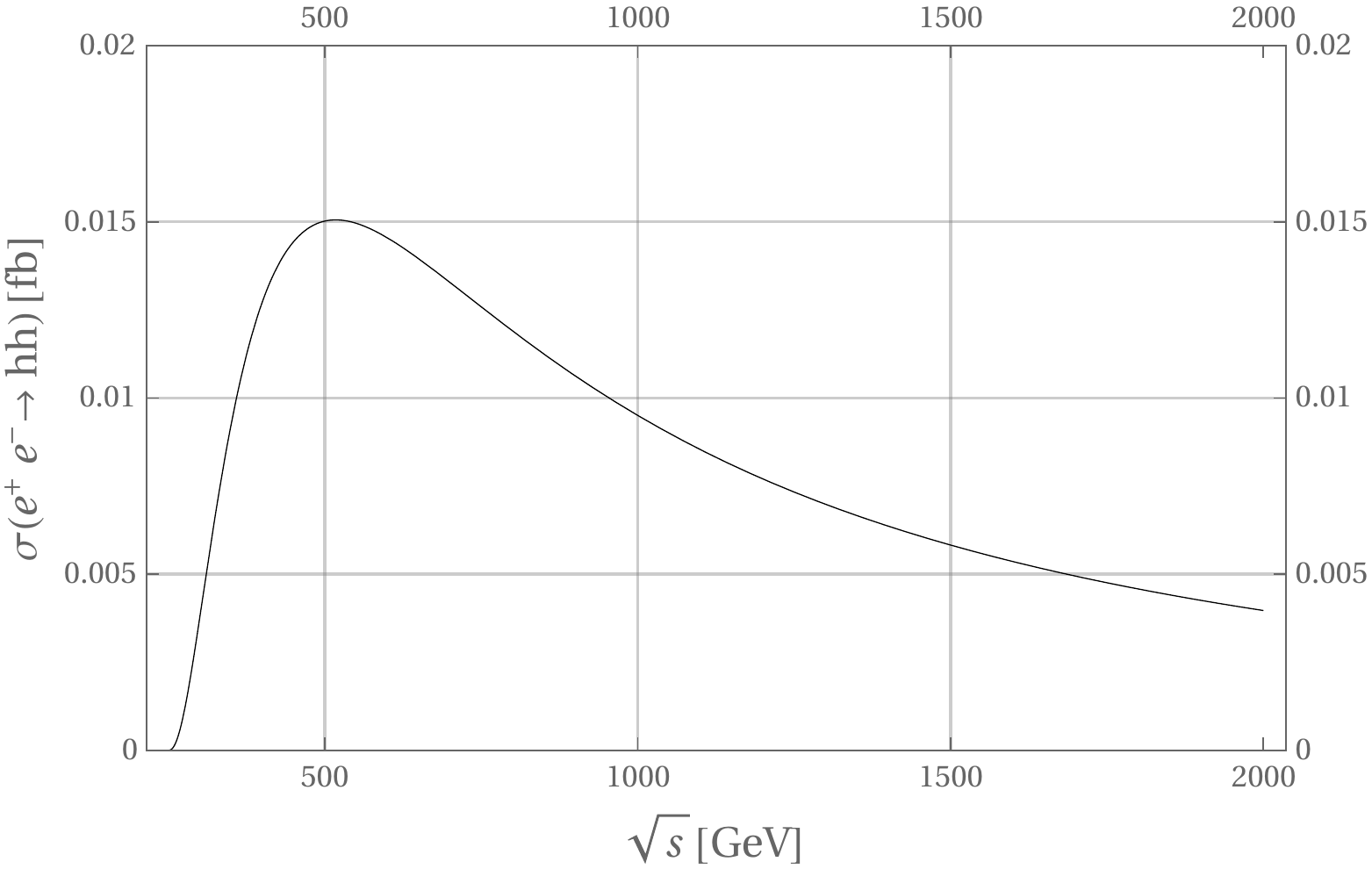} 
\caption{SM cross section for $e^+ e^- \to hh$ as function of the center of mass energy $\sqrt{s}$.} \label{xsection}
\end{figure}

The energy dependence of the leading order SM cross section for $e^+ e^- \to hh$ is shown in Fig.~\ref{xsection}. The cross section acquires its maximum value of approximately $0.015$ fb at around $\sqrt{s}= 500$ GeV.

\section{EFT contributions to $e^+e^-\to hh$}
Double Higgs production at $e^+e^-$ colliders in the SM has been shown to have a tiny cross section of the order of fraction of femtobarns (see Fig.~\ref{xsection}) as discussed in the previous section. However, with large luminosities expected at future $e^+e^-$ colliders, a few hundred events might eventually be collected in the course of a few years, allowing for the experimental study of this final state. On the other hand, cross sections can be enhanced by contributions coming from physics beyond the SM and in this paper we want to entertain this possibility. In particular we will consider effects of new physics parametrized by the presence of higher dimensional operators in the SMEFT framework. The general SMEFT lagrangian can be written as
\be 
{\cal L}_{\rm SMEFT}={\cal L}_{\rm SM}+\sum_{i}\frac{c_i^{(n)}}{\Lambda^{n-4}}{\cal O}_i^{(n)}+\ldots
\ee
where $\Lambda$ is the mass scale of new physics, $c_i^{(n)}$ are dimensionless coefficients and $n$ is the dimension of the gauge invariant operators ${\cal O}_i^{(n)}$ built up with SM fields. It allows for a systematic study of deviations from the SM while respecting established symmetry principles.

In this work we focus on the contributions of dimension-six operators of the SMEFT  because they give the leading contributions in the systematic expansion $E/\Lambda$, where $E$ is the typical energy of the process (the unique dimension-five operator does not contribute to the process $e^+e^-\to h h$). In this work we use the parametrization of~\cite{Grzadkowski:2010es}. In principle, all dimension-six operators that are relevant for the electron and Higgs sector should be considered. However, several of these operators are already constrained from other observables and therefore will not be taken into account in this study. In particular, dimension-six operators that modify the $\bar eeZ$, $e\nu W$, $hZZ$ and $hWW$ vertices are already (strongly) constrained by electroweak precision data and LHC Higgs measurements~\cite{Han:2004az,deBlas:2016ojx,deBlas:2017wmn,Ellis:2014jta,Englert:2015hrx,Butter:2016cvz,Ellis:2018gqa} and we will safely ignore their effects. We are then left with two classes of effective operators that can give sizable contributions: operators that induce an effective $\bar e e h h $ coupling and operators that generate an effective $\bar e e \bar t t$ coupling. The first class enters at tree-level while the second class operators only contribute at one-loop.

There is a unique operator belonging to the first class
\be
\frac{c_{e\varphi}}{\Lambda^2}(\varphi^\dagger \varphi-\frac{\upsilon^2}{2})\bar l_L\varphi e_R +\rm{h.c.}
\label{op1}
\ee  
On the other hand there are seven four-fermion operators belonging to the second class, however six of them give zero contribution because of their chirality structure and in the end we are left with just one four-fermion operator
\be 
\frac{c_{et}}{\Lambda^2}\epsilon_{ij}\bar l_L^ie_R\bar q_L^jt_R +\rm{h.c.}
\label{op2}
\ee
In the equations above $c_{e\varphi}$ and $c_{et}$ 
are dimensionless coefficients, $\Lambda$ is the scale of new physics, $l=(\nu \,\,e)$, $q=(t \,\,b)$, $\varphi$ is the Higgs doublet and $\epsilon_{ij}$ is the total antisymmetric tensor of rank 2.

The operator in Eq.~\eqref{op1} has been written with the constant piece $\upsilon^2/2$ subtracted to the invariant $\varphi^\dagger \varphi$ term in order to formally maintain the tree level relation $m_e=y_e\upsilon/\sqrt{2}$ also in the effective theory. This mass relation is however altered by the potentially sizable loop correction to the electron mass coming from the top-quark loop induced by the effective operator in Eq.~\eqref{op2}. The contribution of this effective operator to the electron self energy in dimensional regularization is given by
\begin{fmffile}{selfelectron}
\setlength{\unitlength}{0.7mm}
\begin{equation}
  \begin{gathered}
	\begin{fmfgraph*}(30,30)
	\fmfset{thin}{0.02pt}
	\fmfset{arrow_len}{2mm}
	\fmfleft{i}
	\fmfright{o} 
	\fmf{fermion,label=$t$,l.side=right,tension=0.7}{v,v}
	\fmf{fermion}{i,v,o}
	\fmffreeze
	\fmfv{label=\textcolor{blue}{$c_{et}$}, l.angle=-90, l.dist=5thick, d.fill=full, fore=(0,,0,,150),d.shape=circle,d.size=7}{v}
	\fmflabel{$e^-$}{i}
	\fmflabel{$e^+$}{o}
	\end{fmfgraph*}
  \end{gathered}
   \qquad \quad = \quad -i\Sigma_e \quad =  \quad  -i\frac{6}{\left(4\pi\right)^2} \frac{c_{et}}{\Lambda^2} m_t^3 \left(1+\frac{1}{\bar{\epsilon}}+\log\frac{\mu^2}{m_t^2}\right) 
\end{equation}
 \end{fmffile}
where $1/\bar \epsilon =1/\epsilon-\gamma+\log 4\pi$. Thus the inverse electron propagator reads
\be 
\slashed{p}-y_{e}\frac{\upsilon}{\sqrt{2}}-\delta y_{e}\frac{\upsilon}{\sqrt{2}}-\Sigma_{e}
\ee
In $\overline{\rm MS}$ the Yukawa counterterm is chosen to be
\be \label{counteryuk}
\delta y_{e}=-\frac{6}{(4\pi)^{2}}\frac{\sqrt{2}}{\upsilon}\frac{c_{et}}{\Lambda^2} m_t^3 \frac{1}{\bar \epsilon}
\ee
such that the physical electron mass is given by\footnote{If one had used an on-shell scheme, then the electron mass definition would have remained unchanged while the Yukawa counterterm would have been modified including also the finite and $\mu$-dependent piece. In the end the two schemes give the same result, as it should be.}
\be \label{electronmass}
m_{e}=y_{e}\frac{\upsilon}{\sqrt{2}}+\frac{6}{(4\pi)^{2}}\frac{c_{et}}{\Lambda^2}m_{t}^{3}\left(1+\log\frac{\mu^{2}}{m_{t}^{2}}\right)
\ee
From the theoretical point of view this mass correction may introduce a fine tuning problem and in order to avoid it one must require that $|\delta m_e| \lesssim m_e$. In this case we have that
\be \label{finetuning}
\left|\frac{c_{et}}{\Lambda^2}\right|\lesssim \frac{8\pi^{2}}{3}\frac{m_e}{m_t^3}\simeq 2\times 10^{-3} {\rm TeV}^{-2}
\ee
By inverting Eq.~\eqref{electronmass} it is possible to express the relation between the Yukawa coupling and the $c_{et}$ coefficient as follows
\be \label{yukred}
y_{e}(\mu)=\frac{\sqrt{2}}{\upsilon}m_{e}-\frac{6}{(4\pi)^{2}}\frac{\sqrt{2}}{\upsilon}\frac{c_{et}}{\Lambda^2}m_{t}^{3}\left(1+\log\frac{\mu^{2}}{m_{t}^{2}}\right)
\ee 
Therefore, thanks to this relation, tree level diagrams of Fig. \ref{smtree} proportional to $y_e$  are not negligible anymore if $c_{et}\neq 0$. Notice from eq. \eqref{yukred} that, contrary to the SM case, the limit of vanishing electron mass does not imply a vanishing Yukawa coupling. The scale $\mu$ entering in Eq. \eqref{yukred} will be set equal to $2m_h$ in the computation of $e^+e^-\to hh$. 

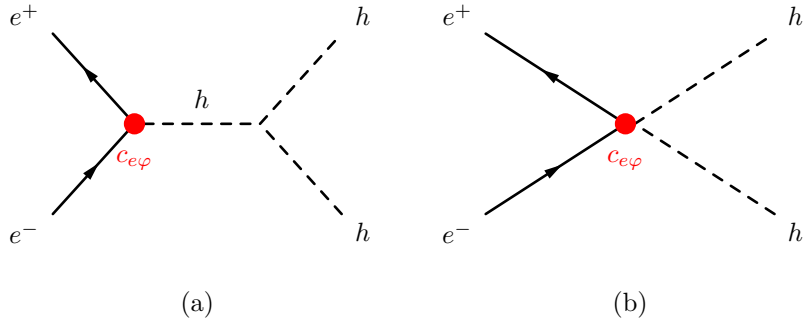
\begin{figure}[ht!]
\vspace{0.9cm}
\unitlength = 1.2mm
\centering
\begin{subfigure}{0.3\textwidth}
\begin{fmffile}{TrLvlEFT2}
\begin{fmfgraph*}(40,20)
\fmfset{thin}{0.8pt}
\fmfset{arrow_len}{2.5mm}
\fmfleft{i1,i2}
\fmfright{o1,o2}
\fmf{fermion}{i1,v1,i2}
\fmf{dashes}{o1,v2,o2}
\fmf{dashes,label=$h$,tension=1.3}{v1,v2}
\fmfv{label=\textcolor{red}{$c_{e\varphi}$},l.angle=-90,l.dist=5thick,f=(1,,0,,0),d.shape=circle,d.size=7}{v1}
\fmflabel{$e^-$}{i1}
\fmflabel{$e^+$}{i2}
\fmflabel{$h$}{o1}
\fmflabel{$h$}{o2}
\end{fmfgraph*}
\end{fmffile}
\label{TrLvlEFT2}
\caption{ }
\end{subfigure}
\qquad
\begin{subfigure}{0.3\textwidth}
\begin{fmffile}{TrLvlEFT}
\begin{fmfgraph*}(40,20)
\fmfset{thin}{0.8pt}
\fmfset{arrow_len}{2.5mm}
\fmfleft{i1,i2}
\fmfright{o1,o2}
\fmf{fermion}{i1,v1,i2}
\fmf{dashes}{o1,v2,o2}
\fmf{dashes,tension=30}{v1,v2}
\fmfv{label=\textcolor{red}{$c_{e\varphi}$},l.angle=-90,l.dist=5thick,f=(1,,0,,0),d.shape=circle,d.size=7}{v1}
\fmflabel{$e^-$}{i1}
\fmflabel{$e^+$}{i2}
\fmflabel{$h$}{o1}
\fmflabel{$h$}{o2}
\end{fmfgraph*}
\end{fmffile}
\label{TrLvlEFT}
\caption{ }
\end{subfigure}
\caption{Tree level contribution to $e^+ e^- \to hh$ coming from Eq.~\eqref{op1}.}
\label{nptree1}
\end{figure}

\begin{figure}[ht!]
\vspace{0.9cm}
\unitlength = 1.2mm
\centering
\begin{subfigure}{0.3\textwidth}
\begin{fmffile}{Loop1EFT}
\begin{fmfgraph*}(40,20)
\fmfset{thin}{0.8pt}
\fmfset{arrow_len}{2.5mm}
\fmfleft{o1,o2}
\fmfright{i1,i2}
\fmf{dashes,tension=1.4}{v1,i1}
\fmf{fermion,label=$t$,l.side=left}{v1,v3}
\fmf{dashes,tension=1.4}{v2,i2}
\fmf{fermion,label=$t$,l.side=left}{v4,v2}
\fmf{fermion,label=$t$,l.side=left,tension=-0.2}{v2,v1}
\fmf{fermion,tension=1.2}{o1,v3}
\fmf{fermion,tension=1.2}{v4,o2}
\fmf{fermion,tension=40}{v3,v4}
\fmfv{label=\textcolor{blue}{$c_{et}$},l.angle=-90,l.dist=5thick,f=(0,,0,,1),d.shape=circle,d.size=7}{v3}
\fmflabel{$e^-$}{o1}
\fmflabel{$e^+$}{o2}
\fmflabel{$h$}{i1}
\fmflabel{$h$}{i2}
\end{fmfgraph*}
\end{fmffile}
\label{Loop1EFT}
\caption{ }
\end{subfigure}
\qquad
\begin{subfigure}{0.3\textwidth}
\begin{fmffile}{Loop2EFT}
\begin{fmfgraph*}(40,20)
\fmfset{thin}{0.8pt}
\fmfset{arrow_len}{2.5mm}
\fmfleft{i1,i2}
\fmfright{o1,o2}
\fmf{fermion}{i1,v2,i2}
\fmf{fermion,left=0.8,tension=0.2,label=$t$}{v3,v2}
\fmf{fermion,left=0.8,label=$t$}{v2,v3}
\fmf{dashes,label=$h$}{v3,v4}
\fmf{dashes}{o1,v4,o2}
\fmfv{label=\textcolor{blue}{$c_{et}$},l.dist=5thick,f=(0,,0,,1),d.shape=circle,d.size=7}{v2}
\fmflabel{$e^-$}{i1}
\fmflabel{$e^+$}{i2}
\fmflabel{$h$}{o1}
\fmflabel{$h$}{o2}
\end{fmfgraph*}
\end{fmffile}
\label{Loop2EFT}
\caption{ }
\end{subfigure}
\caption{Loop level contributions $e^+ e^- \to hh$ coming from Eq.~\eqref{op2}}
\label{nploop1}
\end{figure}
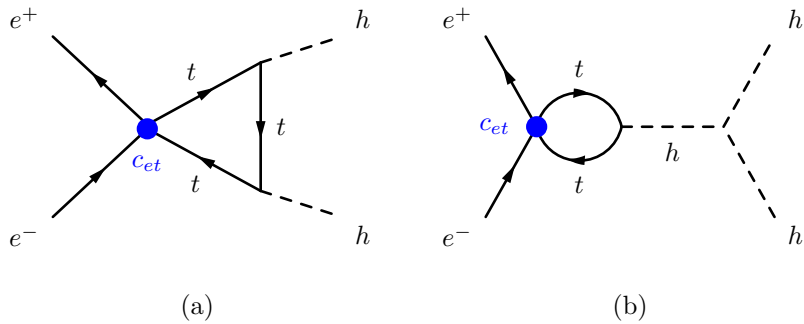

The operator in Eq.~\eqref{op1} introduces a tree level coupling of the electron to the Higgs given by 
\be 
g_{\bar e e h}=\frac{c_{e\varphi}\upsilon^2}{\Lambda^2\sqrt{2}}
\ee 
After considering all contributions to the $\bar e e h$ vertex, it is possible to show that the recent upper bound on the electron Yukawa coupling obtained from Higgs decay~\cite{Altmannshofer:2015qra}  $y_e<600\, y_{e}^{\rm SM}$ implies that
\be \label{hdecaybound}
\left|-\frac{m_e}{\upsilon}+\frac{c_{e\varphi}(\mu)\upsilon^2}{\Lambda^2\sqrt{2}}-\frac{3}{(4\pi)^{2}}\frac{y_t}{\sqrt{2}}\frac{c_{et}}{\Lambda^2} (4m_{t}^{2}-m_h^2)\left[f(m_h^2,m_t^2)+\log\frac{\mu^{2}}{m_{t}^{2}}\right]\right|\lesssim 600 \frac{m_e}{\upsilon}
\ee
where $f(m_h^2,m_t^2)$ is given in Appendix A. The operator in Eq.~\eqref{op1}, besides modifying the $\bar e e h$ vertex, induces also an effective $\bar e e hh$ coupling given by
\be 
g_{\bar eehh}=\frac{3c_{e\varphi}\upsilon}{2\Lambda^2\sqrt{2}}
\ee
which is not present in the SM. This operator contributes at tree level to $e^+e^-\to hh$, as shown in Fig.~\ref{nptree1}.
On the other hand, the operator in Eq.~\eqref{op2} contributes  to $e^+e^-\to hh$ through the counterterm related to the redefinition of the Yukawa coupling of eq.~\eqref{yukred} and it also enters directly at one loop, as shown by the diagrams of Fig.~\ref{nploop1}.

Notice that the operator in Eq.~\eqref{op1} plays also the role of the counterterm needed to absorb the divergence produced by the one-loop insertion of the operator in Eq.~\eqref{op2} and its coefficient $c_{e\varphi}$ has to be formally taken as function of the renormalization scale $\mu$. For the explicit derivation of the counterterm see Appendix B. Therefore, in the process we are studying the coefficients $c_{e\varphi}$ and $c_{et}$ are both formally evaluated at the scale $\mu=2m_h$. 

In our computation we consider just the leading contributions of the operator of Eq.~\eqref{op1} which arise at tree level while the contributions of the operator of Eq.~\eqref{op2} comes at one loop. The total cross section turns out to be a pure quadratic function of the coefficients $c_{e\varphi}$, $c_{et}$, namely the only sizable new physics contributions are of order  $c_{e\varphi}^2$, $c_{et}^2$ and $c_{e\varphi}c_{et}$ because linear terms coming from the interference between SM diagrams, which are helicity conserving,  and new physics diagrams, which are helicity flipping, turns out to be proportional to $m_e/\upsilon$ and therefore negligible. Helicity selection rules and non-interference effects in the context of dimension-six operators have been studied in~\cite{Azatov:2016sqh}. Notice that the EFT expansion is under control because possible interference terms expected from dimension-eight operators which would give comparable contribution in term of the $1/\Lambda^4$ expansion are proportional to $m_e/\upsilon$ as well.

\section{Analysis and results}
We compute the $e^+e^-\to hh$ cross section~$\sigma=\sigma(\frac{c_{e\varphi}}{\Lambda^2},\frac{c_{et}}{\Lambda^2})$ as function of the effective couplings as discussed in the previous section. In order to perform this calculation, we first implemented the effective lagrangian in \texttt{FeynRules}~\cite{Alloul:2013bka} and generate the corresponding \texttt{FeynArts} model output. We used \texttt{FeynArts} 3.10~\cite{FeynArts} and \texttt{FormCalc} 8.4~ \cite{FormCalc} to compute the tree and one-loop amplitudes relevant for the process in the chiral limit ($m_e=0$). Finally, we use \texttt{LoopTools}~\cite{Hahn:1998yk} to compute numerically the cross section as a function of the center of mass energy and effective couplings. We further checked the cross section computation by means of the development version of~\texttt{NLOCT}~\cite{Degrande:2014vpa} and \texttt{MadGraph5\_aMC@NLO}~\cite{Alwall:2014hca}. 

In order to extract the expected 95\% CL limits on the effective operators couplings we assume the measured cross section to coincide with the SM predictions and we construct the following $\chi^{2}$ function
\begin{equation}
\chi^{2}=\chi^{2}(\frac{c_{e\varphi}}{\Lambda^2},\frac{c_{et}}{\Lambda^2})=\frac{\left[\sigma(\frac{c_{e\varphi}}{\Lambda^2},\frac{c_{et}}{\Lambda^2})-\sigma_{{\rm SM}}\right]^{2}}{\delta\sigma^{2}}
\label{chisq}
\end{equation}
where $\sigma_{{\rm SM}}=\sigma(0,0)$. The total cross section uncertainty $\delta\sigma$ that enters in the $\chi^{2}$ computation is given by the combination of the expected experimental $\delta\sigma_{\rm exp}$ and theoretical uncertainties $\delta\sigma_{\rm th}$. In our analysis we assume the theoretical uncertainty to be negligible such that the total uncertainty coincides with the expected experimental one, namely $\delta\sigma=\delta\sigma_{\rm exp}$, which is given by the sum in quadrature of statistic $\delta\sigma_{\rm stat}$ and systematic uncertainties $\delta\sigma_{\rm sys}$
\be 
\delta\sigma=\sqrt{\delta\sigma_{\rm stat}^2+\delta\sigma_{\rm sys}^2}=\sqrt{\frac{\sigma_{\rm SM}}{{ L}}+\alpha^{2}\sigma^{2}_{\rm SM}}\,.
\ee 
The statistical uncertainty is taken to be $\delta\sigma_{\rm stat}=\sqrt{\sigma_{\rm SM}/L}$, where $L$ is the integrated luminosity. The systematic uncertainty has been parametrized by $\delta\sigma_{\rm sys}=\alpha\, \sigma_{\rm SM}$, in analogy to the study performed in~\cite{Durieux:2017rsg}, where $\alpha$ is a dimensionless coefficient that represents the magnitude of the systematic error in relation to the SM cross section. We take a conservative value and we fix $\alpha=0.1$, which corresponds to a 10\% error. However, the impact of the systematic uncertainty will be marginal since our total uncertainty turns out to be statistics dominated due to the expected smallness of the SM cross sections.

We consider different benchmark values of the center of mass energy and luminosity that have been proposed for the future $e^+ e^-$ machines (see Table \ref{colliders1}) and for each configuration we determine 95\% CL limits on the operator coefficients. Values of the coefficients for which $\chi^{2}>3.84$ are considered excluded.

To perform a more realistic investigation we have to consider a set of possible final states that are assumed to be measured at future $e^+ e^-$ colliders in order to reconstruct the Higgs particle through its decay channels. Once a set of final states $(f \bar f)$ and the corresponding branching ratio ${\rm BR}(h\to f \bar f)$ have been identified, then we need to properly rescale the cross section and uncertainty that enter in the chi-squared function of Eq.~\eqref{chisq} by a factor $k={\rm BR}(h\to f_1 \bar f_1)\times {\rm BR}(h\to f_2 \bar f_2)$. For instance, if we assume that each Higgs particle is going to be reconstructed only through its decay to $b \bar b$ then $k\sim 0.35$. 

\begin{table}[h!]
\begin{center}
\begin{tabular}{|c|c|c|c|c|c|} 
\hline 
Benchmark & Experiment & $\sqrt{s}$ (GeV) & $L$ (ab$^{-1}$)& $|c_{e\varphi}/\Lambda^2|({\rm TeV}^{-2})$ & $|c_{et}/\Lambda^2|({\rm TeV}^{-2})$ \tabularnewline
\hline 
\hline 
1 & FCC-ee & 350  & 2.6  & $\quad <0.003 \,\,\,\, (<0.004)\quad$ & $\quad <0.116 \,\,\,\, (<0.146) \quad $\tabularnewline
\hline 
2 & CLIC & 380 & 0.5  &  $<0.004\,\,\,\, (<0.006)$& $<0.143\,\,\,\, (<0.184)$\tabularnewline
\hline 
3 & ILC & 500  & 4 & $<0.003\,\,\,\, (<0.004)$ & $<0.068\,\,\,\, (<0.083)$\tabularnewline
\hline 
4 & CLIC & 1500  & 1.5  & $<0.003\,\,\,\, (<0.003)$ & $<0.027\,\,\,\, (<0.035)$\tabularnewline
\hline
5 & CLIC & 3000  & 3.0 & $<0.002\,\,\,\, (<0.002)$ & $<0.012\,\,\,\, (<0.015)$\tabularnewline
\hline  
\end{tabular}
\caption{\label{colliders1} Table of the different benchmark scenarios considered in our analysis. Each benchmark consists of a specific value of the center of mass energy $(\sqrt{s})$ and luminosity $(L)$ that has been proposed for the future $e^+ e^-$ colliders.  The last two columns represent the 95 \% CL intervals for each operator coefficient taken individually in the analysis with $k=1$ $(k=0.35)$. } 
\end{center}
\end{table}

\begin{figure}
\includegraphics[scale=0.8]{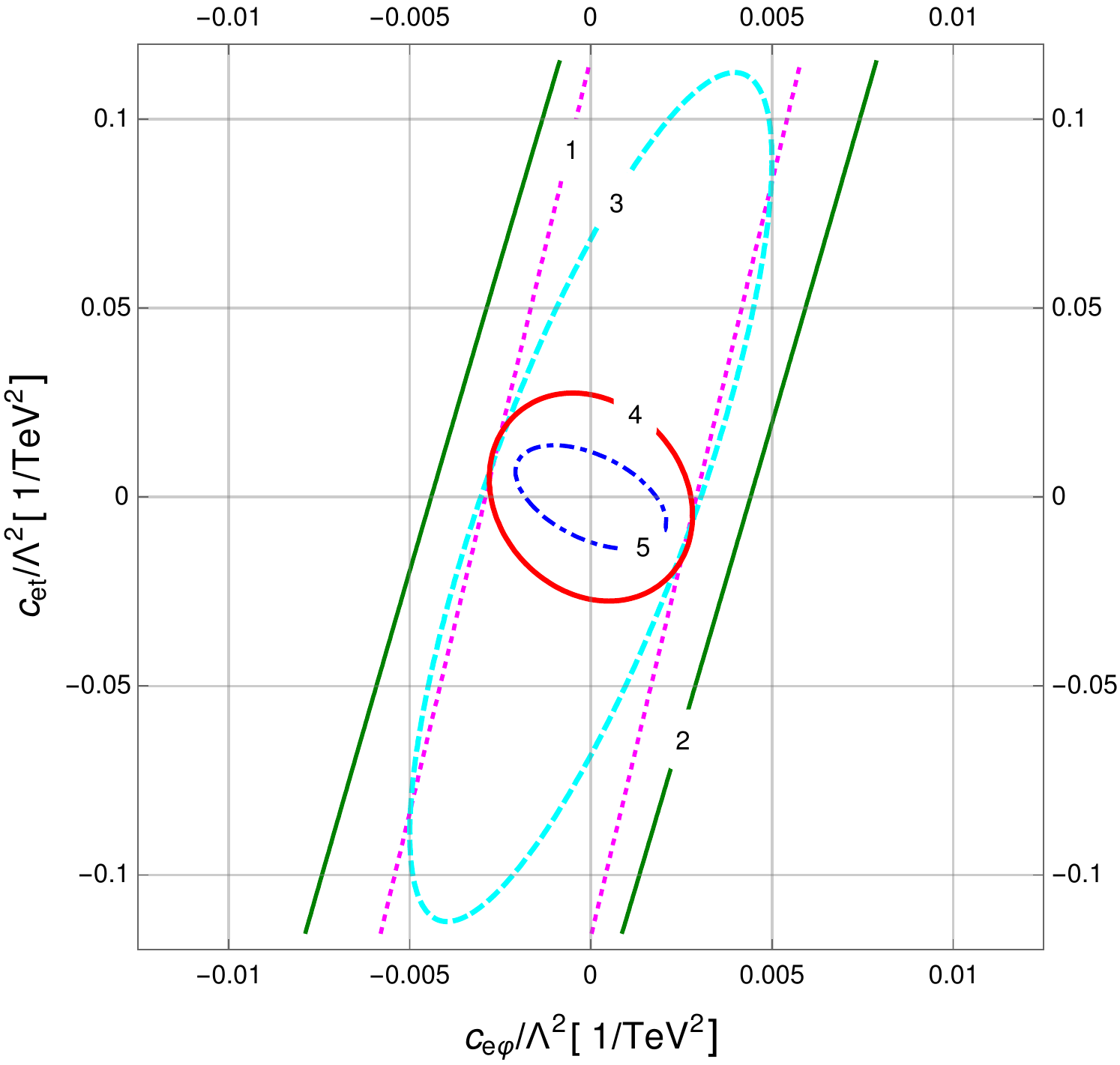} 
\caption{Exclusion regions in the $(c_{e\varphi}/\Lambda^2,c_{et}/\Lambda^2)$ plane for the different benchmark configurations of energy and luminosity reported in Table \ref{colliders1} in the case $k=1$. Points that lie outside the ellipses are excluded at 95\% CL.} \label{Bounds_k10and5}
\end{figure}

The results for $k=1$ $(k=0.35)$ in which each operator is considered individually are reported in the last two columns of Table \ref{colliders1}. The table shows that all benchmark configurations considered in our study provide the same order of magnitude bound for the coefficient $c_{e\varphi}/\Lambda^2$, which is $|c_{e\varphi}/\Lambda^2|\lesssim 3\times 10^{-3}$ TeV$^{-2}$ for $k=1$. This behaviour is expected since the contribution to the total cross section of the operator in Eq.~\eqref{op1}  is almost insensitive to the energy in the process. Assuming an order one coefficient for $c_{e\varphi}$ this implies a quite strong bound on the new physics scale of the order $\Lambda \gtrsim 18$ TeV.

On the other hand, the bound on the coefficient $c_{et}/\Lambda^2$ turns out to be weaker than the bound on $c_{e\varphi}/\Lambda^2$. This is expected since the $c_{et}/\Lambda^2$ contribution enters at one-loop compared to $c_{e\varphi}/\Lambda^2$ which enters at tree level. Moreover, the bound depends on the benchmark configuration considered, because the contribution to the total cross section of the operator in Eq.~\eqref{op2} turns out to be quite sensitive of the energy of the process. Assuming an order one coefficient for $c_{et}$ and $k=1$, the weakest bound $|c_{et}/\Lambda^2|\lesssim 0.15$ TeV$^{-2}$ is obtained from one of the benchmark configurations with lowest center of mass energy and luminosity and can be translated into $\Lambda \gtrsim 2.5$ TeV, while the strongest bound $|c_{et}/\Lambda^2|\lesssim 0.01$ TeV$^{-2}$ is obtained from the benchmark configuration with highest center of mass energy and can be translated into $\Lambda \gtrsim 10$ TeV. The case $k=0.35$ shows modifications of the bounds of the order of 25-50 \% with respect to the $k=1$ case. The fine tuning bound on $c_{et}/\Lambda^2$ in Eq.~\eqref{finetuning} is one order of magnitude stronger than the best expected bound coming from our analysis, however one has to keep in mind that the fine tuning bound is based on theoretical considerations while our bound is based on experimental measurements. Moreover, the actual fine tuning could be milder thanks to cancellations induced by additional operators that we are not considering in our study.

The results for $k=1$ in which both effective operator coefficients are taken into account are shown in Fig.~\ref{Bounds_k10and5}. For each benchmark configuration, the exclusion region is represented by an ellipse. Points that lie outside the ellipse are considered excluded at 95\% CL. By inspection of Fig.~ \ref{Bounds_k10and5}, we can infer that the best sensitivity is given by benchmark scenario number 5 which is characterized by the highest, among the considered configurations, center of mass energy of 3000 GeV. In Fig~\ref{Bounds_k10and5} we do not show the bound of Eq.~\eqref{hdecaybound} coming from Higgs decay since it is weaker than the expected bounds we obtained from $e^+e^-\to h h$ and the exclusion region would lie outside the range of the plot.

The results for $k=0.35$ are not presented since they differ from the results in Fig.~\ref{Bounds_k10and5} by $\sim$30\% and the corresponding ellipses do not present significant modifications.

\section{Conclusions}
Double Higgs production at future $e^+ e^-$ colliders offers the possibility to explore the sensitivity to dimension-6 operators involving electrons that have not been constrained yet. The small SM cross section and the clean environment make this process an ideal laboratory for these studies. In particular, two operators are relevant for this process and are characterized by dimensionless Wilson coefficients   $c_{e\varphi}$ and $c_{et}$. By including their contributions to the double Higgs cross section we derived 95\% bounds based on several benchmarks for these future colliders under certain assumptions of final decay channels to be reconstructed and the errors. We found that the bounds on $c_{e\varphi}$ typically probe scales of ${\cal O} (10\; \mbox{TeV})$ while the $c_{et}$ operator is less constrained since it enters only at one-loop level (of course, more stringent limits on $c_{et}/\Lambda^2$ of ${\cal O}(10^{-3})$ TeV$^{-2}$ can be obtained by studying top quark pair production at future $e^+e^-$ colliders, as shown in~\cite{Durieux:2018tev}). In conclusion, searches for $e^+ e^- \rightarrow hh$ should also be pursued in addition to the more traditional double Higgs production in double higgstrahlung and vector boson fusion in order to explore these possible new couplings.

\section*{Acknowledgments}
AT would like to thank ICTP-SAIFR and IFT-UNESP for hospitality. The work of AV was supported by a CAPES MSc fellowship. RR is partially supported through a CNPq grant 307925/2017-0 and a Fapesp 2016/01343-7 grant. AV and CD are supported by the Fund for Scientific Research F.N.R.S.
through the F.6001.19 convention. We thank E. Pont\'on for a valuable discussion about the correction to the electron mass induced by higher dimensional operators and the renormalization procedure. AV would like to thank F. Maltoni for discussion.

\appendix
\section{$\bar e e h$ coupling modification}
The  effective operators in Eq.~\eqref{op1} and \eqref{op2} modify the $\bar e e h$ coupling with respect to the SM case as follows
\be\label{eehshift}
-\frac{m_e}{\upsilon}\to -\frac{m_e}{\upsilon}+\frac{c_{e\varphi}(\mu)\upsilon^2}{\Lambda^2\sqrt{2}}-\frac{3}{(4\pi)^{2}}\frac{y_t}{\sqrt{2}}\frac{c_{et}}{\Lambda^2} (4m_{t}^{2}-q^2)\left[f(q^2,m_t^2)+\log\frac{\mu^{2}}{m_{t}^{2}}\right]
\ee
where $q$ is the Higgs momentum and
\be
f(q^2,m_t^2)=2+\sqrt{1-\frac{4m_t^2}{q^2}}\log\frac{2m_t^2-q^2+\sqrt{q^2(q^2-4m_t^2)}}{2m_t^2}
\ee
Eq.~\eqref{eehshift} has been obtained by taking into account the tree level contribution to the $\bar e e h$ vertex coming from the $c_{e\varphi}$ operator, the redefinition of the Yukawa coupling in eq.~\eqref{yukred}, the top-loop diagram induced by the $c_{et}$ operator and the proper counterterms.
 
\section{Divergent $e^{+}e^{-}\to hh$ diagrams and counterterms}
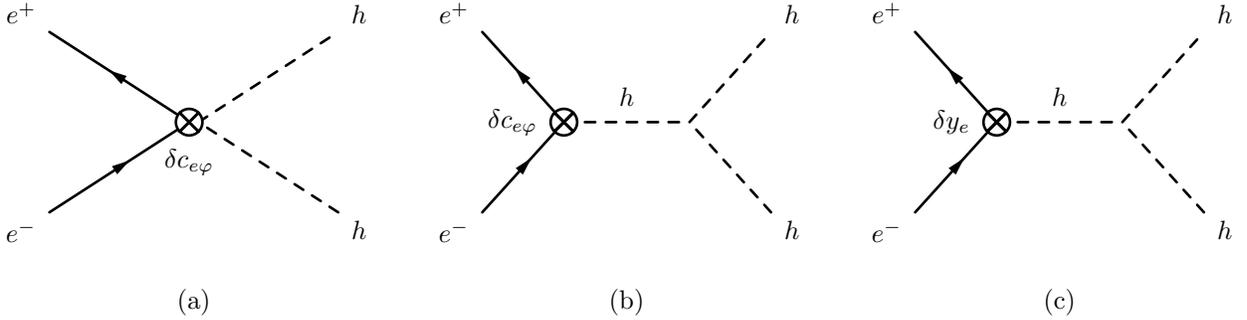
\begin{figure}[ht!]
\vspace{0.9cm}
\unitlength = 1.2mm
\centering
\begin{subfigure}{0.3\textwidth}

\begin{fmffile}{CTerm1}
\begin{fmfgraph*}(40,20)
\fmfcmd{
    path quadrant, q[], otimes;
    quadrant = (0, 0) -- (0.5, 0) & quartercircle & (0, 0.5) -- (0, 0);
    for i=1 upto 4: q[i] = quadrant rotated (45 + 90*i); endfor
    otimes = q[1] & q[2] & q[3] & q[4] -- cycle;
}
\fmfwizard
\fmfset{thin}{0.8pt}
\fmfset{arrow_len}{2.5mm}
\fmfleft{i1,i2}
\fmfright{o1,o2}
\fmf{fermion}{i1,v1,i2}
\fmf{dashes}{o1,v2,o2}
\fmf{dashes,tension=30}{v1,v2}
\fmfv{label={$\delta c_{e\varphi}$},l.angle=-90,d.sh=otimes,l.d=10,d.f=empty,d.size=10}{v1}
\fmflabel{$e^-$}{i1}
\fmflabel{$e^+$}{i2}
\fmflabel{$h$}{o1}
\fmflabel{$h$}{o2}
\end{fmfgraph*}
\end{fmffile}
\label{CT1}
\caption{ }
\end{subfigure}
\qquad
\begin{subfigure}{0.3\textwidth}
\begin{fmffile}{CTerm2}
\begin{fmfgraph*}(40,20)
\fmfcmd{
    path quadrant, q[], otimes;
    quadrant = (0, 0) -- (0.5, 0) & quartercircle & (0, 0.5) -- (0, 0);
    for i=1 upto 4: q[i] = quadrant rotated (45 + 90*i); endfor
    otimes = q[1] & q[2] & q[3] & q[4] -- cycle;
}
\fmfwizard
\fmfset{thin}{0.8pt}
\fmfset{arrow_len}{2.5mm}
\fmfleft{i1,i2}
\fmfright{o1,o2}
\fmf{fermion}{i1,v1,i2}
\fmf{dashes}{o1,v2,o2}
\fmf{dashes,label=$h$,tension=1.3}{v1,v2}
\fmfv{label={$\delta c_{e\varphi}$},l.angle=180,d.sh=otimes,l.d=10,d.f=empty,d.size=10}{v1}
\fmflabel{$e^-$}{i1}
\fmflabel{$e^+$}{i2}
\fmflabel{$h$}{o1}
\fmflabel{$h$}{o2}
\end{fmfgraph*}
\end{fmffile}
\label{CT2}
\caption{ }
\end{subfigure}
\qquad
\begin{subfigure}{0.3\textwidth}
\begin{fmffile}{CTerm3}
\begin{fmfgraph*}(40,20)
\fmfcmd{
    path quadrant, q[], otimes;
    quadrant = (0, 0) -- (0.5, 0) & quartercircle & (0, 0.5) -- (0, 0);
    for i=1 upto 4: q[i] = quadrant rotated (45 + 90*i); endfor
    otimes = q[1] & q[2] & q[3] & q[4] -- cycle;
}
\fmfwizard
\fmfset{thin}{0.8pt}
\fmfset{arrow_len}{2.5mm}
\fmfleft{i1,i2}
\fmfright{o1,o2}
\fmf{fermion}{i1,v1,i2}
\fmf{dashes}{o1,v2,o2}
\fmf{dashes,label=$h$,tension=1.3}{v1,v2}
\fmfv{label={$\delta y_{e}$},l.angle=180,d.sh=otimes,l.d=10,d.f=empty,d.size=10}{v1}
\fmflabel{$e^-$}{i1}
\fmflabel{$e^+$}{i2}
\fmflabel{$h$}{o1}
\fmflabel{$h$}{o2}
\end{fmfgraph*}
\end{fmffile}
\label{CT3}
\caption{ }
\end{subfigure}
\caption{Counterterm diagrams for $e^+ e^- \to hh$.}
\label{CounterTerms}
\end{figure}

The one-loop diagrams (b) and (c) of Fig.~\ref{nploop1} are proportional to $c_{et}$ and UV divergent. The computation in dimensional regularization of the divergent part of these diagrams gives 
\begin{eqnarray}\label{mdiv}
{\cal M}^{(a)div}+{\cal M}^{(b)div} & = & \frac{9}{(4\pi)^{2}}c_{et}2m_{t}\left(y_{t}^{2}-\lambda\right)\left(1+\frac{m_{h}^{2}}{s-m_{h}^{2}}\ \right) \frac{1}{\bar \epsilon}\bar v_e(p_2) u_e(p_1)\nn\\ & & +\frac{9}{(4\pi)^{2}}c_{et}2m_{t}\lambda y_{t}^{2}\frac{\upsilon^{2}}{s-m_{h}^{2}} \frac{1}{\bar \epsilon}\bar v_e(p_2) u_e(p_1)
\end{eqnarray}
where $y_t$ is the top Yukawa, $\lambda$ the Higgs self-coupling, $u_e$ and $v_e$ are the electron and positron Dirac spinors and $1/\bar \epsilon =1/\epsilon-\gamma+\log 4\pi$. Let us now consider the counterterm diagrams proportional to $\delta c_{e\varphi}$ and $\delta y_e$ needed to cancel this divergence, as shown in diagrams (c), (d) and (e) of Fig. \ref{CounterTerms}. We have that
\begin{eqnarray}\label{mcount}
{\cal M}^{(c)ct}+{\cal M}^{(d)ct}+{\cal M}^{(e)ct} & = & -\frac{3\delta c_{e\varphi}}{\sqrt{2}}\upsilon\left(1+\frac{m_{h}^{2}}{s-m_{h}^{2}}\ \right)\bar v_e(p_2) u_e(p_1)\nn\\
&&+\frac{3\delta y_e}{\sqrt{2}\upsilon}\frac{m_{h}^{2}}{s-m_{h}^{2}}\bar v_e(p_2) u_e(p_1)
\end{eqnarray}
By comparing eq. \eqref{mdiv} and \eqref{mcount} we obtain the explicit form of the counterterms in $\overline{\rm MS}$
\begin{equation}
\delta c_{e\varphi}=\frac{6}{(4\pi)^{2}}c_{et}y_{t}\left(y_{t}^{2}-\lambda\right)\frac{1}{\bar \epsilon}
\end{equation}
\begin{equation}\label{counteryukapp}
\delta y_{e}=-\frac{3}{(4\pi)^{2}}c_{et}\upsilon^2y^3_{t}\frac{1}{\bar \epsilon}
\end{equation}
where we have used $m_t=y_t\upsilon/\sqrt{2}$ and $m_h^2=2\lambda \upsilon^2$. Notice that the counterterm in Eq.~\eqref{counteryukapp} coincides with the one derived in Eq.~\eqref{counteryuk}, as it should be. From the explicit form of the counterterm $\delta c_{e\varphi}$ we can read off the contribution of $c_{et}$ to the RG equation of $c_{e\varphi}$ 
\be 
\mu \frac{\partial c_{e\varphi}}{\partial \mu}=\frac{12}{(4\pi)^{2}}c_{et}y_{t}\left(y_{t}^{2}-\lambda\right)
\ee
which agrees with \cite{Jenkins:2013zja} and \cite{Jenkins:2013wua}.

\end{document}